\documentclass[12pt]{iopart}
\usepackage{epsf}
\begin{document}

\title
[Role of electron-phonon interaction in the resistance anisotropy of 2D
electrons]
{On the role of electron-phonon interaction in the resistance anisotropy
of two-dimensional electrons in GaAs heterostructures}

\author{D. V. Fil
\footnote[3]{fil@isc.kharkov.com}
}

\address{Institute for Single Crystals National Academy of Sciences
of Ukraine, Lenin av. 60, Kharkov 61001, Ukraine}

\begin{abstract}
A contribution of the electron-phonon interaction into
the energy of a unidirectional charge ordered state (stripe phase)
of two-dimensional
electrons in GaAs heterostructures is analyzed.
The dependence of the energy on the
direction of  the electron density modulation
is calculated. It is shown that in electrons layers situated close
to the (001) surface
the interference between the
piezoelectric and the deformation potential interaction
causes  a preferential
orientation of the stripes along the [110] axis.
\end{abstract}




\section{Introduction}
The observation of a resistance anisotropy of two-dimensional (2D)
electrons at high Landau level occupancy \cite{1,2,3,4}
is considered as an evidence for a formation of unidirectional
charge ordered states (stripe phases) in such systems.
These states were predicted in Refs. \cite{5,6,7}. The Hartree-Fock
calculations \cite{5,6,7} show that electrons at a topmost
half-filled Landau level are separated in stripes of full and empty
occupancy with a period of about several times the cyclotron radius.
If the stripes are preferentially oriented along a certain direction, such
states are expected to demonstrate the anisotropy in the longitudinal
resistance: the low resistance along the stripes and the high resistance
at the perpendicular direction \cite{8,9}.

The effect is observed in 2D electron layers in GaAs/AlGaAs
heterostructures grown on (001)-oriented GaAs substrates. For
such an experimental setup in the magnetic field perpendicular to the
electron layer the low and high resistance directions are always pinned
along certain crystallography axes of the host matrix, namely,
[110] and [1$\bar{1}$0], correspondingly.

A mechanism that determines a preferential orientation of the stripes
in GaAs heterostructures is not completely understood yet. It was suggested
by Takhtamirov and Volkov \cite{10}, that an anisotropy of the effective mass of 2D
electrons may be responsible for the orientational pinning of the
stripes. In the model developed in Ref. \cite{10} the effective mass
anisotropy is caused by an asymmetry of the quantum well potential confining
the electrons to the (001) plane. This idea has been checked experimentally
by Cooper {\it et al} \cite{11}. In experiments \cite{11} the influence
of the symmetry of the quantum well potential on the resistance anisotropy
has been studied.  It has been found, that the resistance does not depend on
the precise form of the confining potential, in particular, the same
form of the anisotropy is observed for symmetrically
confining 2D electrons.
The authors of Ref. \cite{11} conclude that the asymmetry
of the confining potential cannot be considered as an important factor.

Another possible mechanism for the stripe orientation has been studies by
the present author \cite{12}. It was shown, that the
piezoelectric interaction lowers the
energy of the charge density wave aligned along either the [110] or
[1$\bar{1}$0] axes. Nevertheless, the piezoelectric mechanism alone
does not explain why only one of two preferential orientations is realized.
In this paper we address this problem.
We extend the model proposed in \cite{12} and take into
account, together with the piezoelectric interaction, the
deformation potential interaction. We find that, for the electron
layers situated near the (001) surface of the sample, an
interference between
two  channels of the electron-phonon coupling plays an important role in
the stripe orientation. The mechanism considered explains the resistance
anisotropy observed in the experiments.

\section{The model}

Let us consider a semi-infinite piezoelectric crystal that occupies
a volume $z<0$ and contains an electron layer at the distance $d$
from the surface.
The static energy of the
system can be presented in the form
\begin{equation}
U=\int_{z<0} d^3 r ({{\bf E D}\over 8 \pi }
+{u_{ik} \sigma_{ik}\over 2 })
+ \int_{z>0} d^3 r { E^2\over 8 \pi } + U_{def} \ ,
\label{1}
\end{equation}
where
\begin{equation}
D_i=\varepsilon E_i  - 4 \pi  \beta_{i,kl} u_{kl}
\label{2}
\end{equation}
is the electric displacement field,
\begin{equation}
\sigma_{ik}=\lambda_{iklm} u_{lm} + \beta_{l,ik} E_l \ ,
\label{3}
\end{equation}
the stress tensor,
${\bf E}$, the electric field, $u_{ik}$, the strain tensor,
$\lambda_{iklm}$, the elastic moduli tensor,
$\beta_{i,kl}$, the piezoelectric moduli tensor,
$\varepsilon$, the dielectric constant. To be more specific,
we restrict our consideration to the case  of a cubic lattice.
The last term in (\ref{1}) is the deformation potential interaction.
It is chosen in the form
\begin{equation}
U_{def}=\int d^3 r \Lambda \rho (u_{xx}+u_{yy}) \delta(z+d) \ ,
\label{4}
\end{equation}
where $\Lambda$ is the deformation potential constant, $\rho$,
the 2D electron density. Since we consider the model of an
electron layer of zero-thickness, the interaction with $u_{zz}$ deformations
is not included in (\ref{4}).

The electric and elastic fields in Eq. (\ref{1}) satisfy
the following  equations:
\begin{equation}
\nabla {\bf D} =0 \ ,
\label{5}
\end{equation}
\begin{equation}
{\partial \sigma_{ik}\over \partial x_k }=0
\label{6}
\end{equation}
(at $z<0$), and
\begin{equation}
\nabla {\bf E}=0
\label{7}
\end{equation}
(at $z>0$).

At the free surface ($z=0$) the boundary conditions are
\begin{eqnarray}
D_z \Bigg|_{z=-0}=E_z \Bigg|_{z=+0} \ , \\
E_{x(y)} \Bigg|_{z=-0}=E_{x(y)} \Big|_{z=+0} \ , \\
\sigma_{iz}\Bigg|_{z=-0}=0 \ .
\label{7a}
\end{eqnarray}
At $z=-d$ the normal component of the electric displacement field is
discontinuous
\begin{equation}
D_z \Bigg|_{z=-d+0}-D_z\Bigg|_{z=-d-0} = 4 \pi e \rho \ .
\label{8}
\end{equation}
The deformation potential interaction induces a
tangential force applied to the medium in the $z=-d$ plane.
Rewriting the energy (\ref{4}) in the form
\begin{equation}
U_{def}=-\int d^2 r \sum_{i=x,y} u_i \Lambda  \partial_i \rho \ ,
\label{9}
\end{equation}
where ${\bf u}$ is the displacement vector, we obtain
the following expression for the force,
applied to the unit area:
\begin{equation}
{\bf F}= \Lambda ({\partial \rho\over \partial x},
{\partial \rho\over \partial y},0) \ .
\label{10}
\end{equation}
In equilibrium, this force is compensated by the stresses, and
at $z=-d$
the stress tensor satisfies the boundary condition
\begin{equation}
\sigma_{iz}\Bigg|_{z=-d+0}-\sigma_{iz}\Bigg|_{z=-d-0}=-F_i \ .
\label{11}
\end{equation}

Using equations (\ref{5},\ref{6},\ref{7}) and the boundary conditions
one can reduce the energy (\ref{1}) to the form
\begin{equation}
U={1\over 2}\int d^2 r (e \rho \varphi - u_i F_i) \ ,
\label{12}
\end{equation}
where  the electrostatic potential $\varphi$  and the displacement field
${\bf u}$ are taken
at $z=-d$. Their values are found
from the solution of Eqs. (\ref{5}-\ref{7}) at the
boundary conditions specified.

\section{The stripe state energy anisotropy}

Let us calculate the energy (\ref{12}) for the stripe phase.
We consider the system with the electron density modulated in
a certain direction ${\bf r}_s$. The electron density can be presented as
a Fourier series
\begin{equation}
\rho ({\bf r}_{pl})=\sum_{G_n} e^{i {\bf G}_n {\bf r}_{pl}}
 \rho_{{\bf G}_n} \ ,
\label{13}
\end{equation}
where
$G_n= n {\bf q}$ ($n$ is integer),
${\bf q}\parallel {\bf r}_s$, $|q|= 2\pi /l$, and
$l$ is the period of the stripe structure.
For simplicity, we analyze the case of a unimodal charge density wave
\begin{equation}
\rho ({\bf r}_{pl})=\rho_0 \cos {\bf q r}_{pl} \ .
\label{14}
\end{equation}

We calculate the energy (\ref{12}) as a series in powers of
the electron-phonon interaction constants:
\begin{equation}
\frac{U}{S}=U_0+U_2+\ldots  \ \ ,
\label{15}
\end{equation}
where $S$ is the area of the electron layer,
\begin{equation}
U_0={\pi e^2 \rho_0^2\over 2 q \varepsilon}
\left(1+ {\varepsilon-1\over \varepsilon +1 }e^{-2 q d}\right) \ ,
\label{16}
\end{equation}
the Coulomb energy in the absence the electron-phonon interaction, and
the term $U_2$ is quadratic in the interaction constants.
For concrete physical systems considered below
the contribution of the higher order terms in (\ref{15}) is very small and
it can be neglected.
The term $U_2$ determines the dependence of the energy on the
direction of ${\bf q}$.

Let us first consider an isotropic crystal, for which the sound
velocities do not depend on the direction of the sound propagation.
For a cubic lattice this condition is realized if
the elastic constants satisfy the relation
$c_{11} - c_{12}- 2 c_{44}=0$. For such a special case an analytical
expression for $U_2$ can be presented in a simple form.
For the (001) electron layer the calculation of $U_2$ gives the
following result:
\begin{equation}
U_2= A + B \cos 4 \psi + C \sin 2 \psi \ ,
\label{17}
\end{equation}
where $\psi$ is the angle between ${\bf q}$ and  the [100] axis.
The angle dependence (\ref{17}) is determined by the
parameters $\eta=c_{11}/c_{44}$ and
$\xi=q d$.
Using the strong inequality
$\epsilon\gg 1$, that takes place in $GaAs$, we find for
$B$ and $C$ the following approximate expressions
\begin{equation}
\fl B= E_p\
\Big[ 1-\frac{\eta}{3}   - e^{-2 \xi}
\Big( \frac{(\eta - 3) [ 2 \eta(1-2 \xi) +
\xi^2 (5 \eta- 3)]}
{9(\eta-1)}
+ \xi^3 [\frac{2}{3} (\eta+1)-  \xi
(\eta - 1)]\Big)\Big] \ ,
\label{18}
\end{equation}
\begin{equation}
\fl C=  E_{i}\
e^{-2 \xi}
\Big( {\eta(\eta- 3)(1-\xi)\over \eta-1} +
\xi^2 [ 3\xi (\eta-1)- 2 \eta  - 3] \Big)  \ ,
\label{19}
\end{equation}
where $E_p=
9 \pi^2 e^2 \rho_0^2 e_{14}^2/64 \varepsilon^2 c_{11} q$,
$E_i=\pi |e|\rho_0^2 e_{14} \Lambda /8 \varepsilon c_{11}$.
We do not present here the expression for the parameter $A$ which does
not influence the orientation.

\begin{figure}
\begin{center}
\epsfbox{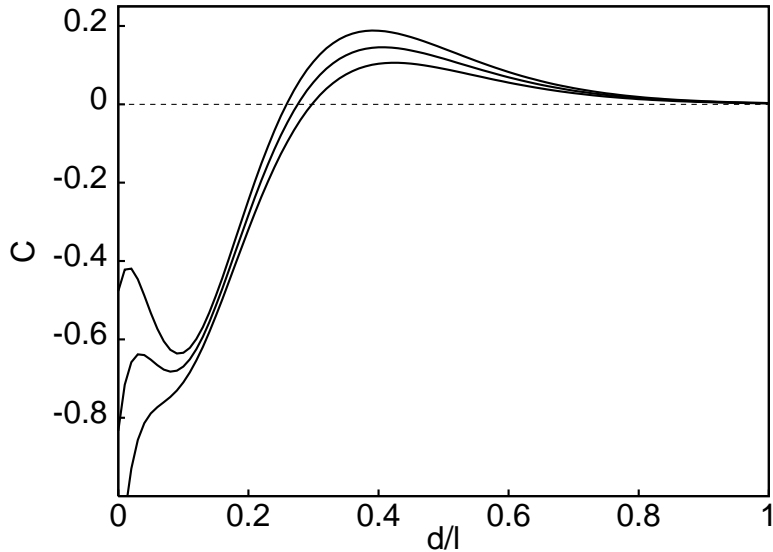}
\end{center}
\caption{\label{f1}The parameter of the stripe energy anisotropy
(Eq.(\ref{19})) in units of $E_i$
for the isotropic crystal; $\eta=2.7,\ 2.5,\ 2.3$
(from  top to bottom).}
\end{figure}

In Eq. (\ref{17}) the second term describes the anisotropy determined
by the piezoelectric interaction, and the third term - the anisotropy
caused by the interference of the piezoelectric and the deformation
potential interaction. One can see that at $\xi\to\infty$
the interference
term tends to zero and the energy (\ref{17}) obeys the $C_{4v}$ symmetry.
If $\xi \sim 1$ the interference term is essential and
the $C_{4v}$ symmetry is reduced to the $C_{2v}$ one.
We find that at $\eta< 2.7$ the
parameter $B$ is positive for all $\xi$ and the global minimum is reached
at  $\psi_m=\pi /4$  or $\psi_m=3\pi /4$ depending on the
sign of the parameter $C$.
At $C<0$ $\psi_m=\pi /4$ and the stripes are preferentially oriented
along the [1$\bar{1}$0] axis, while at $C>0$ $\psi_m=3\pi /4$ and the [110]
oriented stripe phase has the lowest energy.
The dependence of $C$ on $d/l=\xi/2\pi $ is shown in Fig. \ref{f1}.
One can see that at $d/l>0.3$
the [110] orientation of the stripes is realized

\begin{figure}
\begin{center}
\epsfbox{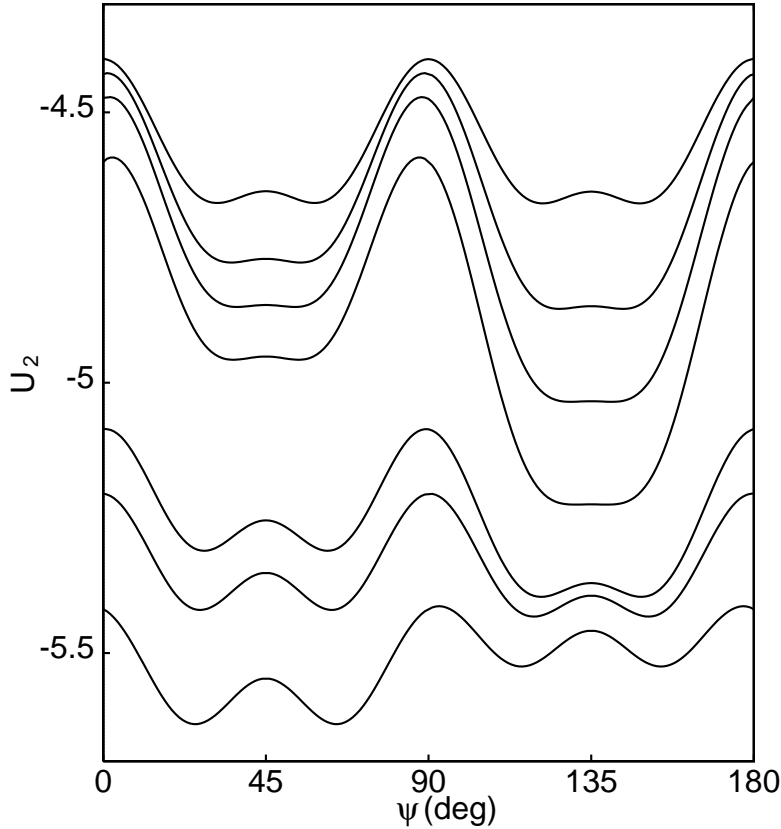}
\end{center}
\caption{\label{f2}The phonon contribution to the energy of the stripe
structure (in units of $E_p$) for GaAs
versus the direction of the electron density modulation
at $d/l=1.0,\ 0.6,\ 0.5,\ 0.4,\ 0.25, 0.23,\  0.2$ (from top to bottom).}
\end{figure}

The results obtained are  sensitive  to the parameters of the
system. Therefore, to determine the stripe
orientation in GaAs heterostructures
it is necessary to take into account the anisotropy of
the elastic moduli. For this case we solve Eqs. (\ref{5}-\ref{7}),
with the boundary conditions  specified,
numerically.
The following parameters
are used for the calculations:
$c_{11}=12.3 \cdot 10^{10}$ $N/m^2$,
$c_{12}=5.4 \cdot 10^{10}$ $N/m^2$,
$c_{44}=6.0 \cdot 10^{10}$ $N/m^2$,
$e_{14}=0.15$ $C/m^2$
$\Lambda=7.4$ $eV$, $\varepsilon =12.5$.
The dependence of $U_2$ on the direction of ${\bf q}$
at $l=2\cdot 10^3$ \AA \ and several $d/l$
is shown
in Fig. \ref{f2}.
In difference with the isotropic crystal, the
minima of the energy are reached at $q$ deviated from the [110] or
[1$\bar{1}$0] axes to the angles $\Delta \psi\approx \pm \pi/12$.
If $d/l$ is in the interval [0.23,1], the configurations with a small
deviation
of $q$ from the [1$\bar{1}$0] axis have the lowest energy. In this case one
can expect that a domain structure is formed, and, in average, the system
should demonstrate the minimum resistance in the [110] direction and the
maximum resistance in the perpendicular direction. The calculations
predict the largest resistance anisotropy at $d/l=0.4$.
At larger and at smaller $d/l$ the anisotropy becomes weaker. At $d/l>1$
and $d/l\approx 0.23$ it  disappears completely. At $d/l<0.23$ the resistance
anisotropy resets, but the high and the low resistance directions
alternates.

\section{Discussion}

We have shown that the classical energy of
a charge density wave in a 2D electron system embedded in a
piezoelectric  matrix
depends on the direction of the wave vector. The effect
is caused by the electron-phonon interaction. The minimum energy is reached
at
two or four different directions of the density modulation. Therefore,
the stripes may form a polydomain structure. In the  bulk samples polydomain
structures may not show any resistance anisotropy, while
in 2D layers situated near the surface of the sample the resistance
anisotropy  should take place.

We calculate the static energy of the stripe structure.
Our results can be also understood as an effect of a virtual exchange of
acoustic phonons between electrons. The electron-electron interaction in
bulk isotropic piezoelectrics
caused by the virtual exchange of acoustic phonons has been studied by
Rashba and Sherman \cite{13}.
Our approach \cite{12} (see  also Ref. \cite{14},
where the
orientation of bi-layer Wigner crystals has been studied)
reproduces the results of Ref. \cite{13}.
Within such an interpretation the effect described in the present paper is
a consequence of
the virtual exchange by surface phonon modes.
For the surface acoustic waves
on the (001) surface
the piezoelectric
and deformation potential interactions give in-phase contributions to the
matrix elements of the electron-phonon interaction \cite{15}, and
the interference between two channel of electron-phonon
interaction takes place.

The unimodal approximation predicts that at $d/l>0.23$
the monodomain stripe structure has the lowest energy for the stripes
deviated from the [110] axis to the angle $\Delta \psi\approx + \pi/12$ or
$\Delta \psi\approx - \pi/12$.
In this case, the polydomain structure
should demonstrate an  anisotropic resistance with a minimum at the [110]
direction.
The largest resistance anisotropy
is reached at $d/l\approx 0.4$.
At $d/l>1$ the anisotropy becomes exponentially weak.
At $d/l<0.23$ the anisotropy changes its sign.
Note, that the last conclusion is specific for the unimodal approximation
(\ref{14}) only. The higher harmonics in (\ref{13}) will shift the
transition point to a smaller value of $d/l$.

We consider our model describes the orientation of stripe structures
formed at high Landau levels. We analyze the anisotropy of the
direct interaction between electrons.
One can expect that the anisotropy of the exchange interaction is  small,
and it does not influence significantly the orientation.

In the Hartree-Fock approximation the period of the electron density
modulation $l$ is approximately equal to
$6 \ell_H$, where $\ell_H$ is the magnetic length. In experiments
\cite{11} the resistance anisotropy was observed at
the  magnetic field $H\approx 2$ T and for $d\approx 2\cdot 10^3$ \AA .
Therefore, for the Hartree-Fork $l$ we find the ratio $d/l\approx 2$.
At such $d/l$ the surface effects are not
important and only an exponentially small violation of the $C_{4v}$ symmetry
may take place.
Since in experiments this violation is quite large
we suppose that the period $l$ is large then it follows
from the Hartree-Fock theory.

It is interesting to evaluate the absolute value of the native anisotropy
caused by the phonon mechanism. Using the parameters given before for
$d/l=0.4$,
the filling factor $\nu =9/2$, the electron density $n=2\cdot 10^{11}$
cm$^{-2}$, and $\rho_0\approx \bar{\rho }$ (where $\bar{\rho }$ is the
average density at the valence Landau level) we find the anisotropy
energy $E_a\approx 0.7$ mK per electron (we determine $E_a$ as the energy
at ${\bf q}\parallel [110]$ minus the energy
at ${\bf q}\parallel [1\bar{1}0]$).

In our consideration we neglect the
screening of the electron-phonon coupling caused by the polarization of
the remote Landau levels. To evaluate the effect of screening in the
unimodal approximation
one can  use
the effective dielectric constant
$\varepsilon_{eff}(q)=\varepsilon (1+K_q)$
and
the screened deformation
potential $\Lambda_{scr}(q)=\Lambda/(1+K_q)$, given by
the random phase approximation. Here
\begin{eqnarray}
\fl K_q=\frac{e^2 q}{\varepsilon \omega_c}
\exp\left(- \frac{q^2\ell_H^2}{2}\right)
\sum_{n,m,\sigma}\!\!\! ^\prime \,
\frac{n!}{m!} \frac{f_{n\sigma} (1-f_{m\sigma })}{m-n}
 \left( \frac{q^2\ell_H^2}{2}\right)^{m-n-1}
\Bigg[L_n^{m-n}\left( \frac{q^2\ell_H^2}{2}\right)\Bigg]^2.
\label{19a}
\end{eqnarray}
In Eq. (\ref{19a}) $f_{n\sigma }=n_F(\varepsilon_{n\sigma })$
is the Fermi factor,
$\omega_c$ is the cyclotron frequency, $L_n^m$ is the generalized Laguerre
polynomial, and the prime on the sum excludes the valence Landau level
(compare with Ref. \cite{16}).
Under
such a substitution,
the form of the dependence  $U_2(\psi)$ (Fig. \ref{f2})
remains almost unchanged,
but the absolute value of the anisotropy is reduced
as $E_a^{src}\approx E_a/(1+K_q)^2$. Evaluation of the formula (\ref{19a})
for the parameters given above yields $E_a^{src}\approx 0.4$ mK.

The phonon contribution to the native anisotropy is comparable
with the one given by the effective mass anisotropy mechanism \cite{10}
(it gives the value of order of 1 mK).
The experimental data for the
anisotropy energy have been obtained in Ref. \cite{11} from the
measurements of the resistivity in  a tilted magnetic field.
The data presented in \cite{11} are based on theoretical
calculations by Jungwirth {\it et al} \cite{16}. According to Ref.\cite{11}
the anisotropy energy for the sample with a conventional heterostructure
is higher (2.4 mK), then for the sample with a symmetric quantum well
(0.5 mK). Thus, we conclude that two mechanisms of the anisotropy work
in parallel, and the survival of the anisotropy in
samples with symmetric quantum wells can
be accounted for the phonon mechanism.

The mechanism of the resistance anisotropy considered is this paper is
essentially dependent on the distance between the surface of the sample
and the electron layer. Therefore, it is desirable to investigate this
dependence experimentally. Such a study may answer the question
whether or not the surface effects play an important role in the stripe
orientation.

It is also of interest to investigate experimentally the influence of the
electron layer orientation on the resistance anisotropy. To
illustrate this point, we outline  the results obtained for the
(111) layer in the isotropic crystal. For such a system the
energy (\ref{17}) is modified to
\begin{equation}
u_{2}= A'+ B' \cos 6 \psi
\label{20}
\end{equation}
where  $\psi$ is counted from the [0$\bar{1}$1] axis. The
coefficient $B'$ is given by the following expression:
\begin{eqnarray}
\fl B'= E_p \frac{10}{27}
\Big( 1- \eta   - \frac{e^{-2 \xi}}{5 (\eta-1)}
[4(\eta^2-\eta-1)(1+2\xi)
+ \xi^2 (\eta+1)(5 \eta-7) \cr  +
\frac{2}{3} \xi^3 (\eta+1)(\eta-5)-
\xi^4 (\eta-1)^2]\Big)
\label{21}
\end{eqnarray}
It is important to note, that the interference term
in (\ref{20}) does not depend on $\psi$ (it is included in $A'$).
We find that the parameter $B'$ is negative for all $\xi$,
and $\eta>2$.
Therefore, the minimum of the energy is reached at
$\psi_m=n \pi/3$ ($n$ is integer).
It means, that monodomain stripe structures should demonstrate
the low resistance along  any of the
[$\bar{2}$11],  [1$\bar{2}$1],or
[11$\bar{2}$] direction, while
the polydomain structures may not show any resistance anisotropy at all.
Since this conclusion is not sensitive to the parameters of the
systems, the lattice anisotropy is not very essential in this case, and
the same behavior is expected for the GaAs system.

\end{document}